\documentclass[12pt]{iopart}
\usepackage{iopams}  
\usepackage{graphicx}
\expandafter\let\csname equation*\endcsname\relax
\expandafter\let\csname endequation*\endcsname\relax
\usepackage{physics}
\usepackage[dvipsnames]{xcolor}
\usepackage{dsfont}
\usepackage[hidelinks]{hyperref}
\usepackage[normalem]{ulem}
\usepackage[utf8]{inputenc}
\usepackage[T1]{fontenc}

\usepackage[english]{babel}
\usepackage{amssymb,amsmath}

\begin{document}

\title{Quantum control and quantum speed limits in supersymmetric potentials}

\author{C.~Campbell$^1$, J.~Li$^2$, Th.~Busch$^1$, and T.~Fogarty$^1$}

\address{$^1$Quantum Systems Unit, Okinawa Institute of Science and Technology Graduate University, Onna, Okinawa 904-0495, Japan}
\address{$^2$Department of Physics, University College Cork, Cork, Ireland}

\ead{christopher.campbell@oist.jp}
    
\vspace{10pt}

\begin{abstract}
Supersymmetry allows one to build a hierarchy of Hamiltonians that share the same spectral properties and which are pairwise connected through common super-potentials. The iso-spectral properties of these Hamiltonians imply that the dynamics and therefore control of different eigenstates are connected through supersymmetric intertwining relations. In this work we explore how this enables one to study general dynamics, shortcuts to adiabaticity (STA) and quantum speed limits for distinct states of different supersymmetric partner potentials by using the infinite box as an example. 
\end{abstract}

\section{Introduction}

Ultracold atoms have become a leading contender for applications in future quantum technologies due to the possibility to control nearly all their degrees of  freedom with high fidelities \cite{RoadmapQuantumOptics,RoadmapAtomtronics}. For their center-of-mass degree of freedom this is due to  immense technical advancements in recent years where almost arbitrary trapping potentials can be designed with spatial light modulators, while optical tweezer arrays allow for the precise control of single particles and the ability to build many-body systems one particle at a time \cite{Nogrette2014,Barredo2018,Gauthier:2021}. One of the few bottlenecks that remains is to design dynamical protocols that allow for minimal operation times of quantum devices, while preventing non-equillibrium excitations from destroying the fragile quantum states and therefore resulting in low fidelity processes. Bounds on this minimal time are commonly known as \textit{quantum speed limits} (QSLs) \cite{Mandelstam:1945}, and are often based on time-energy uncertainty relations. They have been thoroughly researched in recent years and also been extended to mixed states, driven dynamics and open systems \cite{Deffner2017,deCampo2013:open,Oconnor2021,Deffner2021PRX}. 

One class of technique that allows people to develop protocols that operate as close to the QSL as possible while retaining high fidelities are shortcuts to adiabaticity (STAs). These ensure robust adiabatic-like dynamics of quantum states on non-adiabatic timescales and a plethora of techniques to design them are known (see \cite{Odelin:STAreview} for a recent review). Among them are, for example, inverse engineering of optimal time-dependent parameters of a Hamiltonian from the knowledge of the desired adiabatic dynamics of the system \cite{delcampo2011,DeCampo2012:box}, or adding auxiliary fields designed to minimize unwanted excitations during driven dynamics \cite{Berry_2009}. An important subset of STAs deal with scale-invariant dynamics \cite{Deffner2014} which allow to generalize control parameters for different potentials, a well known example being the modulation of the trap frequency of a harmonic oscillator. In fact, while methods in this subset are usually restricted to single-particle systems, in the specific case of the harmonic oscillator they can also be applied to the many-body system of a unitary Fermi gas \cite{Diao_2018} and the self-similar dynamics of the total density can also allow for the experimental extraction of the QSL \cite{Adolfo2021,Garcia2022}.   
The speedup that can be gained from STAs is considered to be of importance for creating quantum devices that suffer from decoherence, but they have recently also found important applications in the area of quantum thermodynamic devices and in particular quantum engines \cite{Abah_2017,Li_2018,Keller2020,Fogarty_2020,Hartmann2020,Myers2020_devicereview}. 

In this work we we explore the dynamics and the use of STAs for the specific situation of a family of Hamiltonians which are related via consecutive supersymmetric transformations \cite{Sukumar_1985,Cooper_1989,Campo2014srep,Zelaya_2020}.
Since the energy spectra of Hamiltonians that are related via a supersymmetric algebra are very similar \cite{ANDRIANOV1984}, this approach allows us to gain insight into the effects on the QSL stemming from the energy for different states in the spectrum vs. the ones coming from the distance between the initial and the final state in Hilbert space. In particular we consider the example of an expanding infinite box and the related higher order supersymmetric partner Hamiltonians \cite{Gutierrez2018,ChrisC2022} and show analytically that the knowledge of the shortcut for the initial Hamiltonian translates through the superpotential to all higher order ones. This also has consequences for the energetic cost of performing the STA and the associated QSL of maintaining the high fidelity dynamics, and which we show can be related through supersymmetric transformations.

\begin{figure}[tb]
    \centering
    \includegraphics[width = \linewidth]{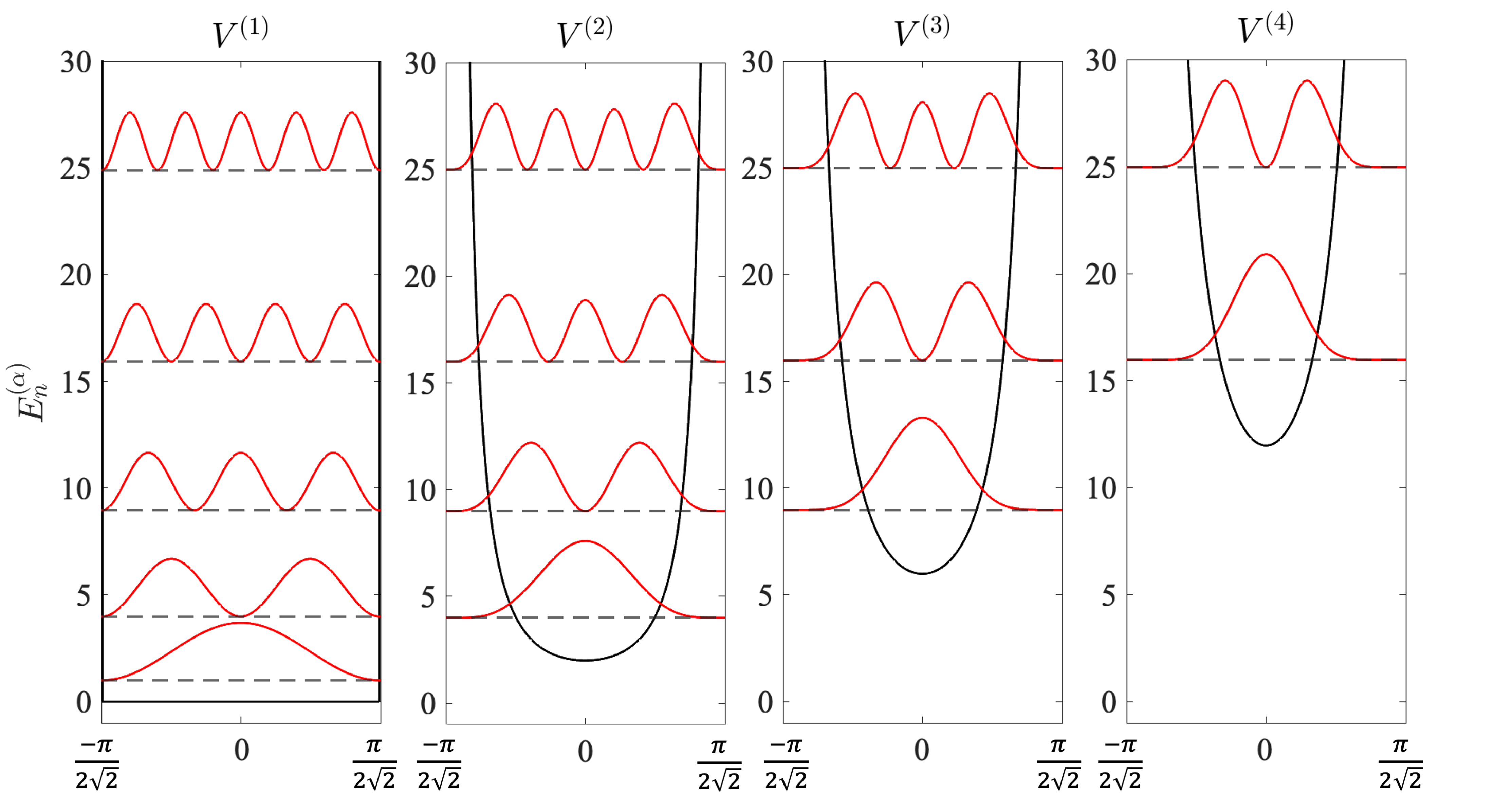}
    \caption{Eigenspectrum (dashed lines) and eigenfunctions (red lines) of the infinite box potential $V^{(1)}$ and the three next higher order supersymmetric partner potentials (full lines). For the purpose of this work we use $L_i = \frac{\pi}{\sqrt{2}}$. One can see that all eigenspectra are degenerate, however each subsequent potential removes the lowest lying energy of the previous one. }
    \label{fig:Schematic}
\end{figure}

\section{Scale Invariant Dynamics}

Throughout this work we will consider a family of one-dimensional Hamiltonians for a particle in different potentials, which are all parametrized by a length $L(t)$ that can be smoothly changed over a total period $\tau$. The time dependent Hamiltonians are labeled by a positive integer $\alpha$ and given by
\begin{equation}
    H^{(\alpha)}(x,t) = -\frac{\hbar^2}{2m}\frac{\partial^2}{\partial x^2}+V^{(\alpha)}(x,t),
    \label{eq:SPHamiltonian}
\end{equation}
and as our starting point ($\alpha=1$) we choose an infinite box potential given by
\begin{equation}
  V^{(1)}(x,t)=\begin{cases}
    \infty, & \text{if}\; |x|>\frac{L(t)}{2}\;,\\
    0, & \text{if}\; |x|<\frac{L(t)}{2}\;,
  \end{cases}
  \label{eq:box}
\end{equation}
with the well known instantaneous eigenstates $\psi^{(1)}_n(x,t) = \sqrt{\frac{2}{L(t)}}\sin(n\pi\left(\frac{x}{L(t)}+\frac{1}{2}\right))$ and energies $\mathcal{E}_n^{(1)}(t)=\frac{n^2\pi^2\hbar^2}{2mL(t)^2}$.
In this case $L(t)$ gives the width of the box and we parameterize it as
 $L(t)=\gamma(t)L_i$, where $\gamma(t)$ is a time dependent scaling factor,
and we choose it to be a \textit{smoother step function} \cite{perlin2002improving}
\begin{equation}\label{eq:sm_ramp}
    \gamma(t) = \frac{L(t)}{L_i}=1+\left[\frac{L_f}{L_i} - 1\right]\left(\frac{t}{\tau}\right)^3\left[10 +3\frac{t}{\tau}\left(2\frac{t}{\tau}-5\right)\right]\;,
\end{equation}
where $L_i$ is the length at $t=0$ and $L_f$ at the final time $\tau$. This ramp for the infinite box potential has previously been studied \cite{DeCampo2012:box} and in the following we will explore the dynamics induced by this ramp for a hierarchy of Hamiltonians which are related to $H^{(1)}$ via a supersymmetric algebra and therefore have an almost identical energy spectrum.

In its simplest form, the algebra of supersymmetric quantum mechanics looks to create pairs of Hamiltonians using a set of unique adjoint operators, namely $H^{(\alpha)} = A^{(\alpha)\dagger} A^{(\alpha)}+E^{(\alpha)}_1$ and $H^{(\alpha+1)} = A^{(\alpha)}A^{(\alpha)\dagger}+E^{(\alpha)}_1$, where  $\alpha \ge 1$ accounts for the order of the supersymmetric transformation \cite{Sukumar_1985,Cooper_1989,COOPER:SUSYBIBLE:1997}. These operators act as creation and annihilation operators and can be explicitly written as 
\begin{align}
    A^{(\alpha)}=& \frac{\hbar}{\sqrt{2m}}\frac{d}{dx} + \mathcal{W}^{(\alpha)}(x)\;,\\
    A^{(\alpha)\dagger} =& -\frac{\hbar}{\sqrt{2m}}\frac{d}{dx} + \mathcal{W}^{(\alpha)}(x)\;,
    \label{eq:SUSY_ops}
\end{align} 
where $\mathcal{W}^{(\alpha)}(x)$ is the so-called superpotential. This allows us to write the partner potentials of Hamiltonians $H^{(\alpha)}$ and $H^{(\alpha+1)}$ as
\begin{equation}\label{potentials}
    V^{(\alpha),(\alpha+1)} = \left[\mathcal{W}^{(\alpha)}(x,t)\right]^2 \mp \frac{\hbar}{\sqrt{2m}}\frac{d}{dx}\left[\mathcal{W}^{(\alpha)}(x,t)\right]\;,
\end{equation}
with the sign between the two terms being negative (positive) for $H^{(\alpha)}$ ($H^{(\alpha+1)}$) . Since in our case $V^{(1)}$ is  the infinite box potential (see Eq.~\eqref{eq:box}), one can find an explicit expression for the superpotential $\mathcal{W}^{(1)}$ as
\begin{eqnarray}
    \mathcal{W}^{(1)}(x,t)&=& -\frac{\hbar}{\sqrt{2m}}\partial_x\ln{(\psi_1^{(1)}(x,t))}\nonumber \\
    &=& \frac{\hbar\pi}{\sqrt{2m}L(t)}
    \tan\left(\frac{\pi x}{L(t)}\right)\;,
\end{eqnarray}
and consequently the form of the first partner potential as
\begin{equation}
    V^{(2)}(x,t) = \frac{\hbar^2\pi^2}{2mL(t)^2}\left(\sec^2\left(\frac{x\pi}{L(t)}\right)
    +\tan^2\left(\frac{\pi x}{L(t)}\right)\right)\;,
\end{equation}
with its ground state \cite{Gutierrez2018,ChrisC2022}
\begin{equation}
    \psi^{(2)}_1 = \sqrt{\frac{2}{3L(t)}}\cos(\frac{\pi x}{L(t)})^2\;.
\end{equation}
This process can be repeated to obtain higher order supersymmetric Hamiltonians and one can find a general form for the superpotential as
\begin{equation}
    \mathcal{W}^{(\alpha)}(x,t) =\frac{\alpha \hbar\pi}{\sqrt{2m}L(t)}
    \tan\left(\frac{\pi x}{L(t)}\right)\;.
\end{equation}
One of the most important features of such a supersymmetric hierarchy of Hamiltonians is that  their eigenspectra are identical, however each higher order one removes the previously lowest lying ground state, i.e. $\mathcal{E}_n^{(\alpha+1)}=\mathcal{E}_{n+1}^{(\alpha)}$ (see Fig.~\ref{fig:Schematic}). The spectrum of any supersymmetric partner Hamiltonian of the infinite square well is therefore given by
\begin{equation}
    \label{eq:heirarchy_energy}
    \mathcal{E}^{(\alpha)}_n(t) = \frac{(n+\alpha-1)^2\pi^2\hbar^2}{2mL(t)^2}\;,
\end{equation}
and one can see the change in the length $L(t)$ effects all spectra in a predictable and straightforward way irrespective of the order of the partner potentials. This allows one to explore the dynamics of different states in the presence of the same spectrum and therefore cleanly separate the influences of the spectrum from the effects stemming from the distance between the initial and final state in Hilbert space. The latter can be quantified by the Bures angle
\begin{equation}
    \mathcal{L}^{(\alpha)} = \arccos{\left(\left|\left\langle \psi_{n}^{(\alpha)}(x,0)|\psi^{(\alpha)}_n(x,\tau)\right\rangle\right|\right)}\;.
\end{equation}
which describes the geometric distance between the initial state at time $t=0$ and the target state at $t=\tau$ \cite{HUBNER1992}.

\begin{figure}[tb]
    \includegraphics[width =\linewidth]{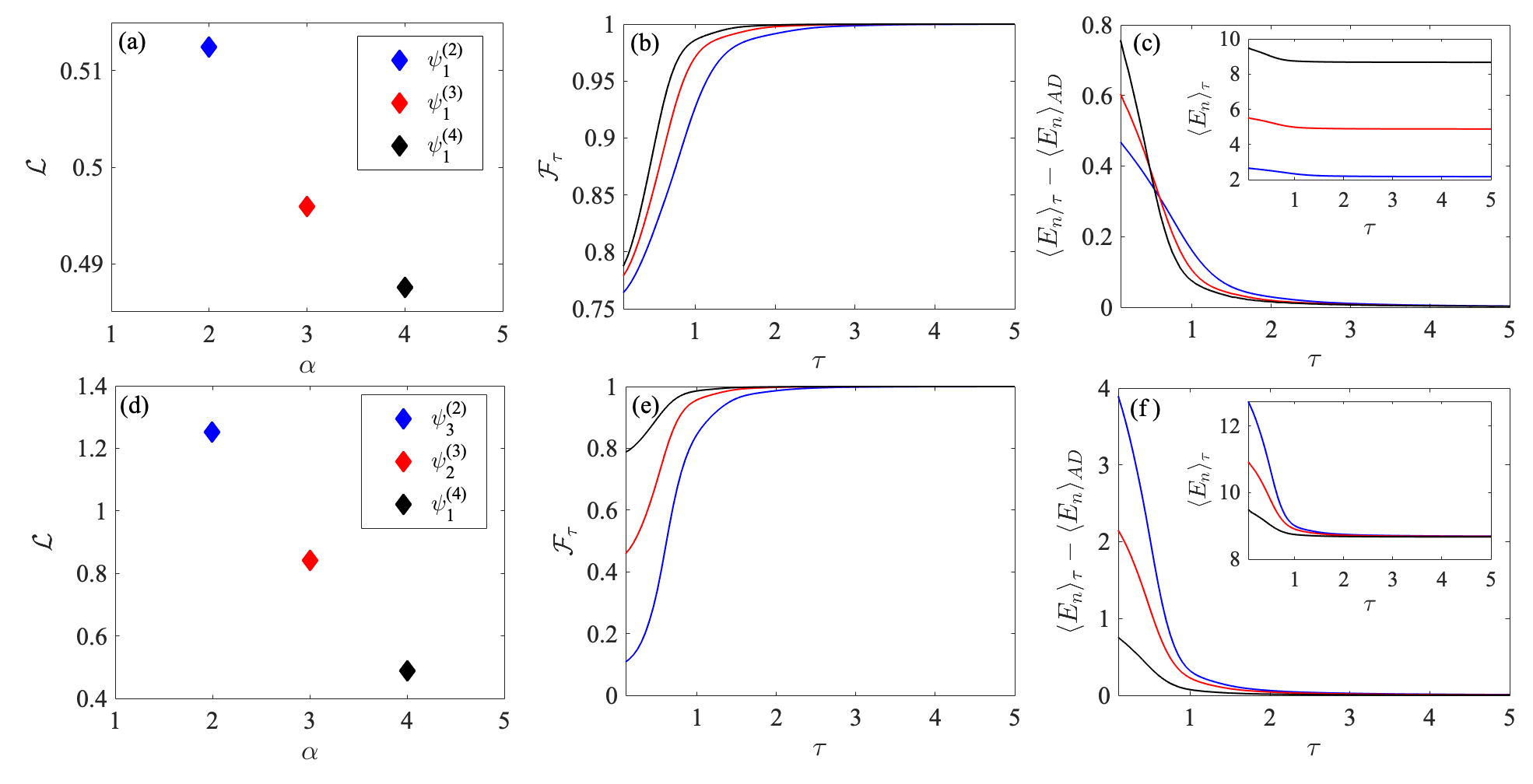}
    \caption{(a,d) Bures angle, (b,e) fidelity and (c,f) time averaged excess energy as a function of the ramp time $\tau$. The insets in panels (c) and (f) show the time averaged energy and we take $L_i = \frac{\pi}{\sqrt{2}}$ throughout. The states are taken from the first 3 partner Hamiltonians of the infinite box with (a-c) all of the ground state wavefunctions and (d-f) wavefunctions with an instantaneous energy of $\mathcal{E}^{(\alpha)}_n(t) = \frac{16\pi^2\hbar^2}{2mL(t)^2}$.}
    \label{fig:bure_fid_en}
\end{figure}

In the following we will study two specific settings in which we double the lengthscale of the initial potential $L_i$, such that at the end of the ramp $L_f=L(\tau) = 2 L_i$. In the first case we look at the respective groundstates $\psi^{(\alpha)}_1$ of the different supersymmetric partner potentials $V^{(2)}, V^{(3)}$ and $V^{(4)}$, for which the Bures angle is shown in Fig.~\ref{fig:bure_fid_en}(a). The Bures angle highlights that the geometric distance that each groundstate needs to travel through Hilbert space is roughly equivalent, whereas the energy of these groundstates, which scales with $\alpha^2$, suggests a different possible speed of evolution for each one. Therefore, the second case we consider is a set of eigenstates that have the same energy within the isospectral setting of different supersymmetric Hamiltonians. In particular we use the states $|\psi_{3}^{(2)}\rangle$, $|\psi_{2}^{(3)}\rangle$ and $|\psi_{1}^{(4)}\rangle$ as the initial state, which all have the same instantaneous energy $\mathcal{E}^{(\alpha)}_n(t) = \frac{16\pi^2\hbar^2}{2mL(t)^2}$. The energy change during an adiabatic ramp is therefore identical for all of them, however, the corresponding Bures angles are quite different (see Fig.~\ref{fig:bure_fid_en}(d)), with the geometric distance growing for increasing quantum number $n$.

Next we explore the finite time dynamics of each initial eigenstate  $\psi^{(\alpha)}_n(x,0)$ during the expansion process using the ramp given in Eq.~\eqref{eq:sm_ramp}, which leads to a state at the end of the unitary dynamics that is given by $\vert\Psi_n^{(\alpha)}(x,\tau)\rangle=\mathcal{T}e^{-\frac{i}{\hbar} \int_{0}^{\tau} H^{(\alpha)}(x,t') dt'}\vert\psi^{(\alpha)}_n(x,0)\rangle$, where $\mathcal{T}$ is the time-ordering operator. The adiabaticity of the driven dynamics can be quantified using the single particle fidelity
$\mathcal{F}(\tau) = |\langle\psi_n^{(\alpha)}(x,\tau)|\Psi^{(\alpha)}_n(x,\tau)\rangle|^2$, which compares the state at the end of the unitary dynamics $\Psi^{(\alpha)}_n(x,\tau)$ with the instantaneous eigenstate $\psi^{(\alpha)}_n(x,\tau)$. If the length scale of the potential is changed slowly, $\tau\rightarrow\infty$, the fidelity is one as $\Psi^{(\alpha)}_n(x,\tau)=\psi^{(\alpha)}_n(x,\tau)$ and the dynamics is considered adiabatic. However, for faster ramps the system can be driven far from equilibrium, which results in a reduced fidelity $\mathcal{F}(\tau)<1$. 

The fidelity for the two sets of initial states we consider is shown as a function of $\tau$ in Figs.~\ref{fig:bure_fid_en}(b) and (e). As expected, in both cases the fidelity is unity for ramp times $\tau\geq4$, while it decreases for faster ramps. However, its is clear that at these short ramp times the fidelity of the two cases is quite different. When  considering the groundstates of each potential (panel (b)) the differences in the fidelity are small, and even for very short times the fidelity is still above 70\%. The reason for this rather high value can be found in the Bures angle, as despite the initial and the target state being different, they are still geometrically rather close in Hilbert space and the system can therefore be driven rather fast. In fact, the fidelity and the Bures angle are similar for all three partner potentials shown here, indicating a  rather similar dynamics. On the other hand, the results for the isospectral initial states possess quite different fidelities in a range from 80\% to 10\%, which is in agreement with the Bures angle indicating that these states have very different geometric distances.

This behaviour is also evident in the amount of non-equilibrium excitations that are created during the trap ramp, which can be quantified using the time average energy
\begin{equation}
    \langle E_n \rangle_\tau = \frac{1}{\tau}\int^{\tau}_{0} dt\; \langle\Psi_n^{(\alpha)}(t)|H^{(\alpha)}(t)|\Psi_n^{(\alpha)}(t)\rangle\;.
\end{equation}
For an adiabatic process, where the evolving state $\Psi_n^{(\alpha)}(t)$ is always an eigenstate of $H^{(\alpha)}(t)$, this average energy is equal to the adiabatic energy $\langle E_n \rangle_{AD} = \frac{1}{\tau}\int^{\tau}_{0} dt\; \mathcal{E}_n^{(\alpha)}(t)$, where the $\mathcal{E}_n^{(\alpha)}(t)$ are the instantaneous energies given in Eq.~\eqref{eq:heirarchy_energy}. In Fig,~\ref{fig:bure_fid_en}(c) and (f) we plot the energy differences $\langle E_n \rangle_\tau-\langle E_n \rangle_{AD}$, which again vanish for large $\tau$, while being finite for faster ramps. Comparing the two cases we consider, the excess energies for the groundstate dynamics only show minor differences, while for the isospectral initial states the non-equilibrium excitations increase the further the states are apart in Hilbert space. This reflects the results for the fidelity, suggesting that the instantaneous energy of each individual state does not play a major role in reaching a high fidelity target state. 

One can further explore the above dynamics by relating the observed results to the QSL time, which bounds the minimum time to connect the two states $\psi_n(x,0)$ and $\psi_n(x,\tau)$. For our system it is given by \cite{Campbell2017:tradeoff}
\begin{equation}
    t \ge \tau_{QSL} \equiv \frac{\hbar}{2\langle E_n\rangle_\tau}[\text{sin}(\mathcal{L})]^2,
    \label{eq:QSL_NA}
\end{equation}
and it is immediately clear that for the groundstates of the supersymmetric partner potentials the QSL time is only dependent on the average energy $\langle E_n\rangle_\tau$, as the Bures angle is effectively unchanged. On the contrary, for isospectral states the average energy for adiabatic ramps ($\tau\rightarrow\infty$) is fixed regardless of the chosen state, while the Bures angle decreases for increasing $n$. Again, the shared properties of supersymmetric Hamiltonians allow to easily compare the dynamical properties of states across the hierarchy of connected Hamiltonians, while adding further insights into controlled dynamics which will be discussed in the next section.

\section{Shortcuts to adiabaticity and Cost}

 The natural next step is to explore the precise control of the expansion dynamics to ensure perfect fidelity for all ramp times $\tau$. For this task we employ  shortcuts to adiabaticity, and while different methods can be applied to our problem \cite{Odelin:STAreview}, here we only focus on the technique of counterdiabatic driving. For this an auxiliary Hamiltonian is added to the single particle Hamiltonian in Eq.~\eqref{eq:SPHamiltonian}, $H^{(\alpha)}_\text{STA}(t)=i\hbar \hat{U}(t)^{\dagger}\partial_t\hat{U}(t)=H^{(\alpha)}(t)+H^{(\alpha)}_\text{CD}(t)$ which effectively nullifies nonadiabtaic excitations and ensures that the driven state always follows the instantaneous eigenstate $\Psi^{(\alpha)}_n(x,t)=e^{-i\int^t \mathcal{E}_n(t')dt'}\psi^{(\alpha)}_n(x,t)$, realizing adiabatic dynamics for any value of $\tau$. Using the time evolution operator $\hat{U}(t,t'=0)=|\Psi^{(\alpha)}_n(x,t)\rangle\langle\psi^{(\alpha)}_n(x,0)|$ the counterdiabatic driving term is given by \cite{Demirplak2003,Demirplak2005,DeCampo2013} 
\begin{equation}\label{eq:alpha_CD}
    H^{(\alpha)}_{\text{CD},n}(t) = i\hbar|\partial_t\psi^{(\alpha)}_n(t)\rangle\langle \psi^{(\alpha)}_n(t)|\;,
\end{equation}
and since the eigenstates of the full hierarchy of the supersymmetric Hamiltonians related to the infinite box can be written in terms of Chebyshev polynomials of the second kind \cite{Gutierrez2018}, it can be exactly evaluated. However, these general expressions can become unwieldly for excited states in higher order Hamiltonians very quickly, and we therefore restrict ourselves here to an illustrative example for the groundstate and first excited state of $H^{(\alpha)}$, which allows us to contrast both groundstate and isospectral STAs. These are given by 
\begin{align}\label{eq:susygs}
    \psi^{(\alpha)}_1 &= \frac{1}{\sqrt{L(t)}}\left[\frac{\sqrt{\pi }\Gamma(\alpha+1)}{\Gamma(\alpha +\frac{1}{2})}\right]^{\frac{1}{2}} \cos(\frac{x\pi}{L(t)})^{\alpha},\\
    \partial_t \psi^{(\alpha)}_1 &= \psi^{(\alpha)}_1\left[-\frac{1}{2}+\frac{\pi x \alpha}{L(t)}\tan(\frac{\pi x}{L(t)})\right]\frac{\dot{L}(t)}{L(t)},
\end{align}
and

\begin{align}\label{eq:susyfirst}
    \psi^{(\alpha)}_2 &= \frac{1}{\sqrt{L(t)}}\left[\frac{2\sqrt{\pi }\Gamma(\alpha+2)}{\Gamma(\alpha +\frac{1}{2})}\right]^{\frac{1}{2}} \sin(\frac{x\pi}{L(t)})\cos(\frac{x\pi}{L(t)})^{\alpha},\\
    \partial_t \psi_2^{(\alpha)}&=\psi_2^{(\alpha)} \left[-\frac{1}{2}-\frac{\pi x}{L(t)} \tan^{-1}\left( \frac{x\pi }{L} \right) + \frac{ x \pi\alpha}{L(t)} \tan\left( \frac{x\pi }{L} \right) \right]\frac{\dot{L}(t)}{L(t)},
\end{align}
and when inserting them into Eq.~\eqref{eq:alpha_CD} one can immediately note that the CD term in both cases is proportional to $\dot{L}(t)/L(t)$ for all $\alpha$, consistent with known STAs for the infinite box potential \cite{Deffner2014,Christopher2014}.

Furthermore, by recalling the intertwining properties of supersymmetric Hamiltonians, $A^{(\alpha)}H^{(\alpha)} = H^{(\alpha+1)}A^{(\alpha)}$ and $H^{(\alpha)} A^{(\alpha)\dagger} = A^{(\alpha)\dagger} H^{(\alpha+1)}$ \cite{Kuru2001}, we can show a similar relation for the counterdiabatic term \cite{Zelaya_2020}. The above intertwining relations between $H^{(\alpha)}$ and $H^{(\alpha+1)}$ allow one to transform two degenerate sets of stationary states through
\begin{align}
    \label{eq:WFIntertwine}
    |\psi^{(\alpha+1)}_{n-1}\rangle = \frac{A^{(\alpha)}}{\sqrt{\Delta \mathcal{E}^{(\alpha)}_n}}|\psi^{(\alpha)}_{n}\rangle\;,\\
    |\psi^{(\alpha)}_{n}\rangle = \frac{A^{(\alpha)\dagger}}{\sqrt{\Delta \mathcal{E}^{(\alpha)}_n}}|\psi^{(\alpha+1)}_{n-1}\rangle\;,
\end{align}
where $\Delta \mathcal{E}^{(\alpha)}_n=\mathcal{E}_n^{(\alpha)}-\mathcal{E}_1^{(\alpha)}$ is the difference in energy between the $n^{th}$ state and the ground state of $H^{(\alpha)}$. This then directly allows one to write an expression for the intertwining relation of the counterdiabatic term as

\begin{align}
    H^{(\alpha+1)}_{\text{CD},n} =& i\hbar \partial_t |\psi^{(\alpha+1)}_n\rangle \langle \psi^{(\alpha+1)}_n| \nonumber \\
    =&\frac{i\hbar}{\Delta \mathcal{E}^{(\alpha)}_{n+1}} 
    \partial_t \left(A^{(\alpha)}|\psi^{(\alpha)}_{n+1}\rangle\right) \langle \psi^{(\alpha)}_{n+1}|A^{(\alpha)\dagger} \nonumber \\
    =& \frac{1}{\Delta \mathcal{E}^{(\alpha)}_{n+1}} A^{(\alpha)} H^{(\alpha)}_{\text{CD},n+1}A^{(\alpha)\dagger}+
    \frac{i\hbar}{\Delta \mathcal{E}^{(\alpha)}_{n+1}}(\partial_t A^{(\alpha)})|\psi^{(\alpha)}_{n+1}\rangle \langle \psi^{(\alpha)}_{n+1}|A^{(\alpha)\dagger}  \;,  
\end{align}
where $H^{(\alpha)}_{\text{CD},n+1}=i\hbar|\partial_t \psi^{(\alpha)}_{n+1}\rangle \langle \psi^{(\alpha)}_{n+1}|$ is the counterdiabatic Hamiltonian for the state $|\psi^{(\alpha)}_{n+1}\rangle$, which has the same instantaneous energy as $|\psi^{(\alpha+1)}_{n}\rangle$, and the second term arises due to the time-dependence of the supersymmetric operator $A^{(\alpha)}$. While this expression for the intertwining relations of the counterdiabatic Hamiltonian is not as simple as for the system Hamiltonian alone, it shows that the counterdiabatic term for the next higher supersymmetric Hamiltonian can be constructed with only knowledge of the lower lying Hamiltonian and its eigenstates \cite{Zelaya_2020}.

\begin{figure}[tb]
    \includegraphics[width =\linewidth]{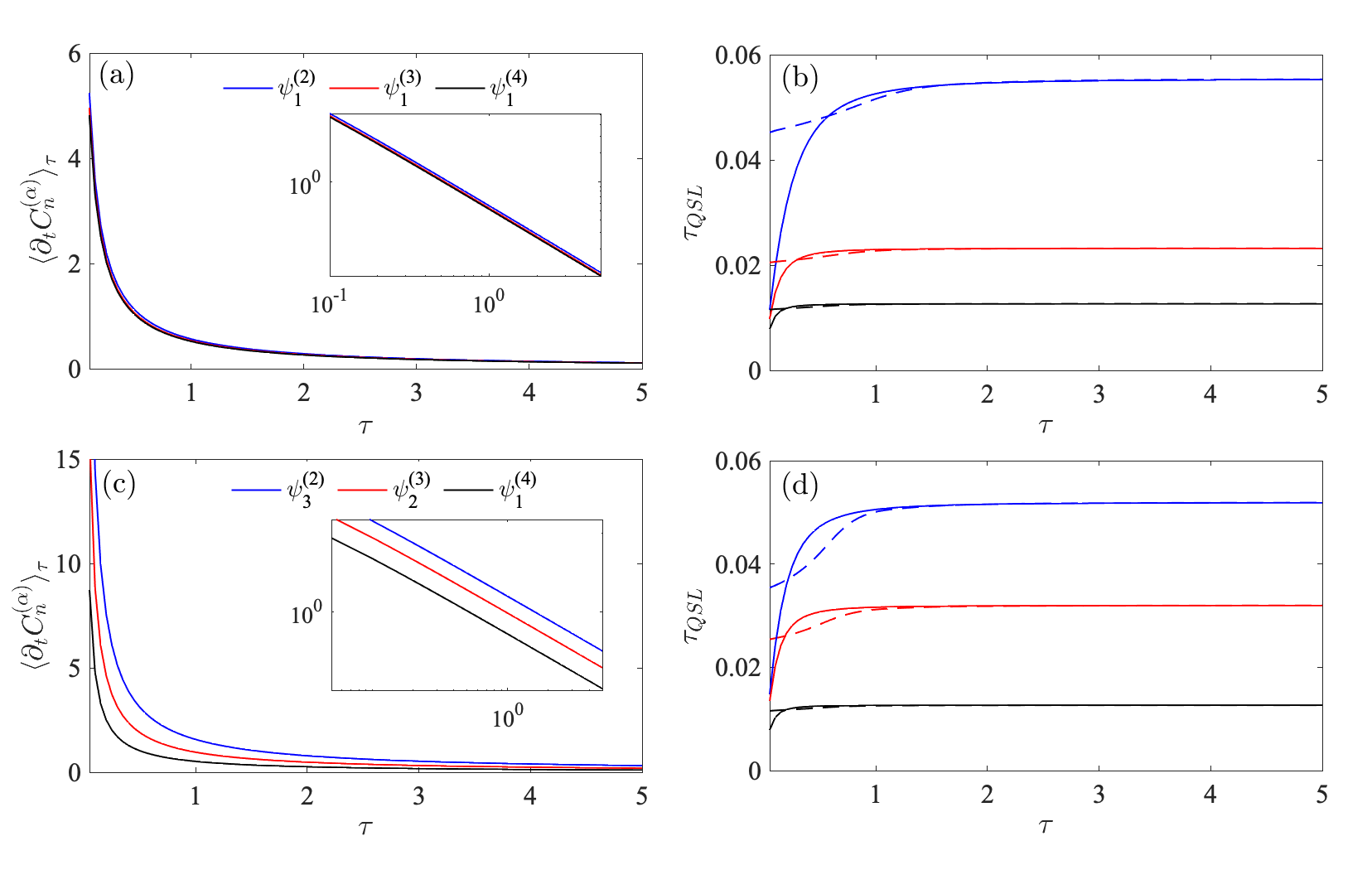}
    \caption{(a,c) Time averaged cost $\langle \partial_t C_n^{(\alpha)}\rangle_{\tau}=\frac{1}{\tau}\int_0^{\tau}\partial_t C_n^{(\alpha)}dt$ (insets are in log-log scale) and (b,d) QSL time as a function of ramp duration $\tau$. Panels (a,b) show results for the ground states of the first three partner Hamiltonians and panels (c,d) for the isospectral states. In panels (b) and (d) the dashed lines are the QSL times of the non-adiabatic dynamics Eq.~\eqref{eq:QSL_NA}, and solid lines are QSL times for the STA dynamics Eq.~\eqref{eq:QSL_STA}.}
    \label{fig:GS_cost_QSL}
\end{figure}
 
Let us next look at the relationship between the quantum speed limit and the energetic cost of the counterdiabatic term \cite{Campbell2017:tradeoff}. In our case this cost is defined as the energy required to achieve adiabatic dynamics for a specific expansion stroke and a common way to quantify this is given by the trace norm of the counterdiabatic driving term $C_n^{(\alpha)} = \int^\tau_0 dt ||H^{(\alpha)}_\text{CD,n}||_{\text{tr}}$ \cite{Zheng2016} where the integrand can be written in terms of the instantaneous eigenstates as
\begin{equation}
    \partial_t C^{(\alpha)}_n = \sqrt{\langle\partial_t\psi^{(\alpha)}_n|\partial_t\psi^{(\alpha)}_n\rangle}\;.
\end{equation}
Taking the above expressions, it is straightforward to find the cost of implementing the STA for the ground state of any supersymmetric Hamiltonian $H^{(\alpha)}$  as
\begin{equation}
    \partial_t C^{(\alpha)}_1 = \frac{\dot{L}(t)}{L(t)}\left[-\frac{1}{4} + \int^{L(t)/2}_{-L(t)/2} \psi_1^{(\alpha)}\psi_1^{(\alpha)*}\left(\frac{\pi x \alpha}{L(t)}\tan(\frac{x\pi}{L(t)})\right)^2 dx\right]^{\frac{1}{2}}\;,
    \label{cost_GS}
\end{equation}
where it is worth noting that the leading term is independent of $\alpha$, while the integral is order dependent. 
To compare isospectral states in different order Hamiltonians the costs can be shown to be again related by supersymmetric transformations. Using the explicit expressions given above for the ground and first excited state of adjacent supersymmetric Hamiltonians we get
\begin{align}
    \partial_t C^{(\alpha-1)}_2 = 
    \frac{\dot{L}(t)}{L(t)}\Biggl[-\frac{1}{4} +
    (2\alpha-1)\int^{L(t)/2}_{-L(t)/2} &\psi_1^{(\alpha)*} \psi_1^{(\alpha)}\left(\frac{x\pi}{L(t)}\right)^2\nonumber\\
    &\times\left(1+(1-\alpha) \tan(\frac{x\pi}{L(t)})^2 \right)^2 dx\Biggr]^{\frac{1}{2}} \;,
\end{align}
with the leading term still independent of $\alpha$. Such a relation can be found for any set of eigenstates in the supersymmetric hierarchy which possess the same instantaneous energy. 

In Fig.~\ref{fig:GS_cost_QSL} we show the time average cost $\langle \partial_t C_n^{(\alpha)}\rangle_{\tau}=\frac{1}{\tau}\int_0^{\tau}\partial_t C_n^{(\alpha)}dt$ as a function of the ramp time $\tau$ for the groundstates (panel (a)). One can see that the cost is diverging for short times  ($\tau\rightarrow0$), where more energy is required to keep the system adiabatic. Interestingly, despite the different states having different energies and also a different energy gap to the next higher state in the spectrum, $\mathcal{E}^{(\alpha)}_2-\mathcal{E}^{(\alpha)}_1 \neq \mathcal{E}^{(\alpha+1)}_2-\mathcal{E}^{(\alpha+1)}_1$, the cost for each groundstate is approximately equal (see inset for log-log scale) as the integral in Eq.~\eqref{cost_GS} only varies weakly with $\alpha$. This is vastly different for the isospectral case shown in panel (c). Here the mean cost increases for higher excited states, which is consistent with the results concerning the fidelty and  energy for the non-adiabatic dynamics shown in Fig.~\ref{fig:bure_fid_en}. We can therefore conclude that states whose wavefunction possesses more nodes  (which have a higher quantum number $n$) are less robust and require a larger energetic cost to remain adiabatic during a controlled expansion.

Finally, in Figs.~\ref{fig:GS_cost_QSL}(b) and (d) we show the QSL time for both the non-adiabatic (Eq.~\eqref{eq:QSL_NA}) and the STA dynamics. For the latter it is given by \cite{Campbell2017:tradeoff}
\begin{equation}
    \tau_{QSL} = \frac{\hbar \tau [\text{sin}(\mathcal{L}_\tau)]^2}{2\int^\tau_0 dt \sqrt{\mathcal{E}^2_n + (\partial_t C)^2}},
    \label{eq:QSL_STA}
\end{equation}
where the cost of the STA is now included in the denominator.
For the groundstate comparison (panel (b)) the geometric distance and the STA costs vary only weakly between the states, but the instantenous eigenvalues $\mathcal{E}_1^{(\alpha)}$ are quite different and are the main factor in the large differences in the QSL time. In comparison, the isospectral states have the same instantaneous energy $\mathcal{E}^{(\alpha)}_n$, however the distance and the cost are unique for each state, which results in the distinct QSL times. Furthermore, we can compare the QSL time for the STA with that of the non-adiabatic driving in Eq.~\eqref{eq:QSL_NA}. Obviously, at long times they converge as both dynamics become adiabatic, while at short times $\tau\rightarrow0$ the STA can lower the QSL time to vanishing values. On the contrary, the non-adiabatic QSL time on these timescales is finite, as for $\tau=0$ this simply describes a sudden quench of the trap potential, while the reduction in the QSL time is a consequence of the non-equilibrium excitations that allow the system to be driven faster from its initial state.

\section{Conclusions}
We have used the specific properties of supersymmetric partner potentials of the infinite box to explore the controlled dynamics through counterdiabatic driving. Supersymmetry provides a framework to explore STA protocols under different conditions and it allows one to compare and contrast different situations while constraining certain parameters. In fact, we have shown that eigenstates of different supersymmetric partner potentials share common properties, such as the Bures angle and eigenspectrum, therefore offering a convenient platform to investigate how these individually contribute to both non-equilibrium dynamics and quantum control. The two sets of eigenstates that we considered in this work, namely the groundstates and a set of isospectral states of each partner potential, are unique to the hierarchy of supersymmetric Hamiltonians and possess distinct dynamics. Specifically, we have shown that both the non-equilibrium and STA dynamics of the groundstates are closely related as they share comparable geometric distances. On the contrary, the dynamics of isospectral states and the degree of control required by the STA, is strongly dependent on the specific state regardless of their energy. Moreover, these states are related through their shared superpotentials which allowed us to show that dynamical quantities, such as the counterdiabatic term and the cost for implementing it, transfer over seamlessly as a consequence of the intertwining properties, allowing one to build the control for the entire hierarchy of Hamiltonians from the knowledge of just one set of known eigenstates. 

Promising extensions to this work could explore dynamically driving between partner Hamiltonians or transforming between isospectral states using time dependent operators $A^{(\alpha)}(t)$ which may be analogous to shortcuts performed by trap deformation \cite{MGaraot2016}. Furthermore, recent experimental advances have shown that supersymmetric potentials can be created using holographic optical traps \cite{Cassettari2022}, which can be used to design unique energy spectra and create isospectral supersymmetric states in the lab.

\section*{Acknowledgements}
 This work was supported by the Okinawa Institute of Science and Technology Graduate University. TF acknowledges support under JSPS KAKENHI-21K13856. The authors are also grateful for the the Scientific Computing and Data Analysis (SCDA) section of the Research Support Division at OIST. J.L. acknowledges support from the Science Foundation Ireland Frontiers for the Future Research Grant ``Shortcut-Enhanced Quantum Thermodynamics" No.19/FFP/6951.

\section*{References}
\bibliographystyle{iopart-num}
\bibliography{biblio}

\providecommand{\newblock}{}
\begin{thebibliography}{10}
\expandafter\ifx\csname url\endcsname\relax
  \def\url#1{{\tt #1}}\fi
\expandafter\ifx\csname urlprefix\endcsname\relax\def\urlprefix{URL }\fi
\providecommand{\eprint}[2][]{\url{#2}}

\bibitem{RoadmapQuantumOptics}
Dumke R, Lu Z, Close J, Robins N, Weis A, Mukherjee M, Birkl G, Hufnagel C,
  Amico L, Boshier M~G, Dieckmann K, Li W and Killian T~C 2016 {\em Journal of
  Optics\/} {\bf 18} 093001
  \urlprefix\url{https://doi.org/10.1088/2040-8978/18/9/093001}

\bibitem{RoadmapAtomtronics}
Amico L, Boshier M, Birkl G, Minguzzi A, Miniatura C, Kwek L~C, Aghamalyan D,
  Ahufinger V, Anderson D, Andrei N, Arnold A~S, Baker M, Bell T~A, Bland T,
  Brantut J~P, Cassettari D, Chetcuti W~J, Chevy F, Citro R, De~Palo S, Dumke
  R, Edwards M, Folman R, Fortagh J, Gardiner S~A, Garraway B~M, Gauthier G,
  G\"unther A, Haug T, Hufnagel C, Keil M, Ireland P, Lebrat M, Li W,
  Longchambon L, Mompart J, Morsch O, Naldesi P, Neely T~W, Olshanii M, Orignac
  E, Pandey S, P\'erez-Obiol A, Perrin H, Piroli L, Polo J, Pritchard A~L,
  Proukakis N~P, Rylands C, Rubinsztein-Dunlop H, Scazza F, Stringari S, Tosto
  F, Trombettoni A, Victorin N, von Klitzing W, Wilkowski D, Xhani K and
  Yakimenko A 2021 {\em AVS Quantum Science\/} {\bf 3} 039201
  \urlprefix\url{https://doi.org/10.1116/5.0026178}

\bibitem{Nogrette2014}
Nogrette F, Labuhn H, Ravets S, Barredo D, B\'eguin L, Vernier A, Lahaye T and
  Browaeys A 2014 {\em Phys. Rev. X\/} {\bf 4}(2) 021034
  \urlprefix\url{https://doi.org/10.1103/PhysRevX.4.021034}

\bibitem{Barredo2018}
Barredo D, Lienhard V, de~L{\'e}s{\'e}leuc S, Lahaye T and Browaeys A 2018 {\em
  Nature\/} {\bf 561} 79
  \urlprefix\url{https://doi.org/10.1038/s41586-018-0450-2}

\bibitem{Gauthier:2021}
Gauthier G, Bell T~A, Stilgoe A~B, Baker M, Rubinsztein-Dunlop H and Neely T~W
  2021 Chapter one - dynamic high-resolution optical trapping of ultracold
  atoms ({\em {Advances In Atomic, Molecular, and Optical Physics}\/} vol~70)
  ed Dimauro L~F, Perrin H and Yelin S~F (Academic Press) pp 1--101
  \urlprefix\url{https://doi.org/10.1016/bs.aamop.2021.04.001}

\bibitem{Mandelstam:1945}
Mandelstam L and Tamm I 1945 {\em J. Phys.\/} {\bf 9} 249

\bibitem{Deffner2017}
Deffner S and Campbell S 2017 {\em Journal of Physics A: Mathematical and
  Theoretical\/} {\bf 50} 453001
  \urlprefix\url{https://doi.org/10.1088/1751-8121/aa86c6}

\bibitem{deCampo2013:open}
del Campo A, Egusquiza I~L, Plenio M~B and Huelga S~F 2013 {\em Phys. Rev.
  Lett.\/} {\bf 110}(5) 050403
  \urlprefix\url{https://link.aps.org/doi/10.1103/PhysRevLett.110.050403}

\bibitem{Oconnor2021}
O'Connor E, Guarnieri G and Campbell S 2021 {\em Phys. Rev. A\/} {\bf 103}(2)
  022210 \urlprefix\url{https://link.aps.org/doi/10.1103/PhysRevA.103.022210}

\bibitem{Deffner2021PRX}
Poggi P~M, Campbell S and Deffner S 2021 {\em PRX Quantum\/} {\bf 2}(4) 040349
  \urlprefix\url{https://link.aps.org/doi/10.1103/PRXQuantum.2.040349}

\bibitem{Odelin:STAreview}
Gu\'ery-Odelin D, Ruschhaupt A, Kiely A, Torrontegui E, Mart\'{\i}nez-Garaot S
  and Muga J~G 2019 {\em Rev. Mod. Phys.\/} {\bf 91}(4) 045001
  \urlprefix\url{https://link.aps.org/doi/10.1103/RevModPhys.91.045001}

\bibitem{delcampo2011}
del Campo A 2011 {\em Phys. Rev. A\/} {\bf 84}(3) 031606
  \urlprefix\url{https://link.aps.org/doi/10.1103/PhysRevA.84.031606}

\bibitem{DeCampo2012:box}
del Campo A and Boshier M~G 2012 {\em Scientific Reports\/} {\bf 2} 648
  \urlprefix\url{https://doi.org/10.1038/srep00648}

\bibitem{Berry_2009}
Berry M~V 2009 {\em Journal of Physics A: Mathematical and Theoretical\/} {\bf
  42} 365303 \urlprefix\url{https://doi.org/10.1088/1751-8113/42/36/365303}

\bibitem{Deffner2014}
Deffner S, Jarzynski C and del Campo A 2014 {\em Phys. Rev. X\/} {\bf 4}(2)
  021013 \urlprefix\url{https://link.aps.org/doi/10.1103/PhysRevX.4.021013}

\bibitem{Diao_2018}
Diao P, Deng S, Li F, Yu S, Chenu A, del Campo A and Wu H 2018 {\em New Journal
  of Physics\/} {\bf 20} 105004
  \urlprefix\url{https://doi.org/10.1088/1367-2630/aae45e}

\bibitem{Adolfo2021}
del Campo A 2021 {\em Phys. Rev. Lett.\/} {\bf 126}(18) 180603
  \urlprefix\url{https://link.aps.org/doi/10.1103/PhysRevLett.126.180603}

\bibitem{Garcia2022}
Garc\'{\i}a-Pintos L~P, Nicholson S~B, Green J~R, del Campo A and Gorshkov A~V
  2022 {\em Phys. Rev. X\/} {\bf 12}(1) 011038
  \urlprefix\url{https://link.aps.org/doi/10.1103/PhysRevX.12.011038}

\bibitem{Abah_2017}
Abah O and Lutz E 2017 {\em {EPL} (Europhysics Letters)\/} {\bf 118} 40005
  \urlprefix\url{https://doi.org/10.1209/0295-5075/118/40005}

\bibitem{Li_2018}
Li J, Fogarty T, Campbell S, Chen X and Busch T 2018 {\em New Journal of
  Physics\/} {\bf 20} 015005
  \urlprefix\url{https://doi.org/10.1088/1367-2630/aa9cd8}

\bibitem{Keller2020}
Keller T, Fogarty T, Li J and Busch T 2020 {\em Phys. Rev. Research\/} {\bf
  2}(3) 033335
  \urlprefix\url{https://link.aps.org/doi/10.1103/PhysRevResearch.2.033335}

\bibitem{Fogarty_2020}
Fogarty T and Busch T 2020 {\em Quantum Science and Technology\/} {\bf 6}
  015003 \urlprefix\url{https://doi.org/10.1088/2058-9565/abbc63}

\bibitem{Hartmann2020}
Hartmann A, Mukherjee V, Niedenzu W and Lechner W 2020 {\em Phys. Rev.
  Research\/} {\bf 2}(2) 023145
  \urlprefix\url{https://link.aps.org/doi/10.1103/PhysRevResearch.2.023145}

\bibitem{Myers2020_devicereview}
Myers N~M, Abah O and Deffner S 2022 {\em AVS Quantum Science\/} {\bf 4} 027101
  \urlprefix\url{https://doi.org/10.1116/5.0083192}

\bibitem{Sukumar_1985}
Sukumar C~V 1985 {\em Journal of Physics A: Mathematical and General\/} {\bf
  18} L57--L61 \urlprefix\url{https://doi.org/10.1088/0305-4470/18/2/001}

\bibitem{Cooper_1989}
Cooper F, Ginocchio J~N and Wipf A 1989 {\em Journal of Physics A: Mathematical
  and General\/} {\bf 22} 3707--3716
  \urlprefix\url{https://doi.org/10.1088/0305-4470/22/17/035}

\bibitem{Campo2014srep}
del Campo A, Boshier M~G and Saxena A 2014 {\em Scientific Reports\/} {\bf 4}
  \urlprefix\url{https://doi.org/10.1038%2Fsrep05274}

\bibitem{Zelaya_2020}
Zelaya K and Hussin V 2020 {\em Journal of Physics A: Mathematical and
  Theoretical\/} {\bf 53} 165301
  \urlprefix\url{https://doi.org/10.1088/1751-8121/ab78d1}

\bibitem{ANDRIANOV1984}
Andrianov A, Borisov N and Ioffe M 1984 {\em Physics Letters A\/} {\bf 105}
  19--22 ISSN 0375-9601
  \urlprefix\url{https://www.sciencedirect.com/science/article/pii/037596018490553X}

\bibitem{Gutierrez2018}
Gutierrez K, Le{\'{o}}n E, Belloni M and Robinett R~W 2018 {\em European
  Journal of Physics\/} {\bf 39} 065404
  \urlprefix\url{https://doi.org/10.1088/1361-6404/aadc7f}

\bibitem{ChrisC2022}
Campbell C, Fogarty T and Busch T 2022
  \urlprefix\url{https://doi.org/10.48550/arXiv.2203.03130}

\bibitem{perlin2002improving}
Perlin K 2002 Improving {N}oise {\em Proceedings of the 29th Annual Conference
  on Computer Graphics and Interactive Techniques\/} SIGGRAPH '02 (New York,
  NY, USA: Association for Computing Machinery) pp 681--682 ISBN 1581135211
  \urlprefix\url{https://doi.org/10.1145/566570.566636}

\bibitem{COOPER:SUSYBIBLE:1997}
Cooper F, Khare A and Sukhatme U 1995 {\em Physics Reports\/} {\bf 251}
  267--385 ISSN 0370-1573
  \urlprefix\url{https://www.sciencedirect.com/science/article/pii/037015739400080M}

\bibitem{HUBNER1992}
Hübner M 1992 {\em Physics Letters A\/} {\bf 163} 239--242 ISSN 0375-9601
  \urlprefix\url{https://www.sciencedirect.com/science/article/pii/037596019291004B}

\bibitem{Campbell2017:tradeoff}
Campbell S and Deffner S 2017 {\em Phys. Rev. Lett.\/} {\bf 118}(10) 100601
  \urlprefix\url{https://link.aps.org/doi/10.1103/PhysRevLett.118.100601}

\bibitem{Demirplak2003}
Demirplak M and Rice S~A 2003 {\em The Journal of Physical Chemistry A\/} {\bf
  107} 9937--9945 \urlprefix\url{https://doi.org/10.1021/jp030708a}

\bibitem{Demirplak2005}
Demirplak M and Rice S~A 2005 {\em The Journal of Physical Chemistry B\/} {\bf
  109} 6838--6844 pMID: 16851769
  \urlprefix\url{https://doi.org/10.1021/jp040647w}

\bibitem{DeCampo2013}
del Campo A 2013 {\em Phys. Rev. Lett.\/} {\bf 111}(10) 100502
  \urlprefix\url{https://link.aps.org/doi/10.1103/PhysRevLett.111.100502}

\bibitem{Christopher2014}
Jarzynski C 2013 {\em Phys. Rev. A\/} {\bf 88}(4) 040101
  \urlprefix\url{https://link.aps.org/doi/10.1103/PhysRevA.88.040101}

\bibitem{Kuru2001}
Kuru {\c{S}}, Te{\u{g}}men A and Ver{\c{c}}in A 2001 {\em Journal of
  Mathematical Physics\/} {\bf 42} 3344--3360
  \urlprefix\url{https://doi.org/10.1063/1.1383787}

\bibitem{Zheng2016}
Zheng Y, Campbell S, De~Chiara G and Poletti D 2016 {\em Phys. Rev. A\/} {\bf
  94}(4) 042132
  \urlprefix\url{https://link.aps.org/doi/10.1103/PhysRevA.94.042132}

\bibitem{MGaraot2016}
Mart\'{\i}nez-Garaot S, Palmero M, Muga J~G and Gu\'ery-Odelin D 2016 {\em
  Phys. Rev. A\/} {\bf 94}(6) 063418
  \urlprefix\url{https://link.aps.org/doi/10.1103/PhysRevA.94.063418}

\bibitem{Cassettari2022}
Cassettari D, Mussardo G and Trombettoni A 2022  (\textit{Preprint}
  \eprint{https://doi.org/10.48550/arXiv.2202.03446})

\end{thebibliography}

\end{document}